\pdfoutput=1

\RequirePackage{fix-cm}

\documentclass[twocolumn]{svjour3}

\smartqed

\usepackage{amsmath}
\usepackage{graphicx,amssymb}
\usepackage{epsf,psfig}
\usepackage{txfonts}
\usepackage{natbib}
\usepackage[unicode]{hyperref}

\journalname{Astrophysics and Space Science}

\begin{document}

\title{Fractal basins of attraction in the planar circular restricted three-body problem with oblateness and radiation pressure}

\author{Euaggelos E. Zotos}

\institute{Department of Physics, School of Science, \\
Aristotle University of Thessaloniki, \\
GR-541 24, Thessaloniki, Greece\\
Corresponding author's email: {evzotos@physics.auth.gr}}

\date{Received: 27 February 2016 / Accepted: 25 April 2016 / Published online: 4 May 2016}

\titlerunning{Fractal basins of attraction in the planar circular restricted three-body problem with oblateness and radiation pressure}

\authorrunning{Euaggelos E. Zotos}

\maketitle

\begin{abstract}

In this paper we use the planar circular restricted three-body problem where one of the primary bodies is an oblate spheroid or an emitter of radiation in order to determine the basins of attraction associated with the equilibrium points. The evolution of the position of the five Lagrange points is monitored when the values of the mass ratio $\mu$, the oblateness coefficient $A_1$, and the radiation pressure factor $q$ vary in predefined intervals. The regions on the configuration $(x,y)$ plane occupied by the basins of attraction are revealed using the multivariate version of the Newton-Raphson method. The correlations between the basins of convergence of the equilibrium points and the corresponding number of iterations needed in order to obtain the desired accuracy are also illustrated. We conduct a thorough and systematic numerical investigation demonstrating how the dynamical quantities $\mu$, $A_1$, and $q$ influence the basins of attractions. Our results suggest that the mass ratio and the radiation pressure factor are the most influential parameters, while on the other hand the structure of the basins of convergence are much less affected by the oblateness coefficient.

\keywords{Restricted three body-problem; Equilibrium points; Basins of attraction}

\end{abstract}

\section{Introduction}
\label{intro}

One of the most interesting and important topics in celestial mechanics as well as in dynamical astronomy is, without any doubt, the classical problem of the circular restricted three-body problem (RTBP). This problem describes the motion of a test particle with an infinitesimal mass under the gravitational field of two primary bodies which move in circular orbits around their common center of gravity \citep{S67}. The applications of this problem expand in many fields of research from chaos theory and molecular physics to planetary physics, stellar systems or even to galactic dynamics. This justifies why this topic still remains active and stimulating.

Over the years, several modifications of the RTBP have been proposed, especially for investigating the character of motion of massless particles in the Solar System. All these modifications include additional types of forces in the total potential function of the classical RTBP in an attempt to take into consideration more dynamical parameters of the physical system and therefore make the study of the motion of the test particle more realistic.

In the classical version of the RTBP the two primary bodies are assumed to be spherically symmetric. In our Solar System however, several celestial bodies, such as Saturn and Jupiter, have found to be sufficiently oblate \citep{BPC99}. The oblateness of a celestial body should be taken into account so as the dynamical exploration of the particular planetary system to be more realistic. The influence of the oblateness coefficient has been studied in a series of papers \citep[e.g.,][]{AEL12,BS12,KMP05,KDP06,KPP08,MPP96,MRVK00,PPK12,SSR75,SSR76,SSR79,SSR86,SRS88,SRS97,SL12,SL13,Z15a,Z15b}.

Another interesting perturbing case is the scenario of a test particle moving in the neighborhood of a radiating primary under the combined influence of radiation and gravitational forces. This problem is known as the photogravitational RTBP. A characteristic example is the motion of a dust grain in the vicinity of a binary stellar system in which one ore even both primary bodies (stars) are emitting radiation thus exerting light pressure to the dust grain. The role of the radiation pressure on the motion of the test particle has been investigated by several authors \citep[e.g.,][]{BC79,DNMY09,KP78,KT85,KPR06,KPP08,KT85,L88,MRVK00,P06,P76,RZ88a,RZ88b,S80,S82,SMB85,T94,Z15c}.

Recently, new additional types of perturbations have been introduced to the classical RTBP. In \citet{B12} and \citet{BHF13} a three-body interaction was added to the inverse-square pairwise gravitational forces in the RTBP. The contribution of this additional force is assumed to inversely depend on the product of the distances of the test particle from the two primaries. This new model, also known as modified circular restricted three-body problem (MCR3BP), is based on the assumption that in a
binary star system containing a sufficient small companion the total gravitational field may not be accurately modeled by the classical pairwise gravitational interactions only. Another interesting non-gravitational perturbation is the Yarkovsky effect \citep[e.g.,][]{E12}.The anisotropic emission of thermal photons, which carry momentum create a force that act on a rotating body \citep[e.g.,][]{R54}. The influence of the Yarkovsky effect may be relatively small however, it is very important especially in celestial mechanics when calculating the proper orbits of small celestial bodies, such as asteroids. Furthermore, when the regime of the rotation of an asteroid changes we have the case of the generalized Yarkovsky effect, i.e. the Yarkovsky-O'Keefe-Radzievskii-Paddack effect or the YORP effect \citep[e.g.,][]{R00}.

In dynamical systems an issue of paramount importance is the determination of the basins of attraction for the equilibrium points (which act as attractors) using an iterative scheme. In other words the sets of initial conditions $(x_0,y_0)$ on the configuration plane which lead to a specific equilibrium point define the several attraction regions (known also as basins of convergence). \citet{D10} used the Newton-Raphson iterative method in order to investigate the basins of attraction in the Hill's problem with oblateness and radiation pressure. In the same vein, the multivariate version of the same iterative scheme has been used to unveil the basins of convergence in the restricted three-body problem \citep[e.g.,][]{KGK12}, the four-body problem \citep[e.g.,][]{BP11,KK14}, or even the ring problem of $N + 1$ bodies \citep[e.g.,][]{CK07}. In the present study we shall try to determine the Newton-Raphson basins of attraction in the RTBP with oblateness and radiation pressure.

The structure of the paper is as follows: In Section \ref{mod} we describe the basic properties of the considered mathematical model. In section \ref{lgevol} the evolution of the position of the equilibrium points is investigated as the values of the main dynamical quantities of the system vary in predefined intervals. In the following Section, we conduct a thorough numerical exploration revealing the Newton-Raphson basins of attraction and how they are affected by the mass ratio, the oblateness coefficient and the radiation pressure factor. Our paper ends with Section \ref{disc}, where the discussion and the conclusions of this work are presented.

\section{Description of the mathematical model}
\label{mod}

According to the classical circular restricted three-body problem (RTBP), the two primary bodies, $P_1$ and $P_2$ with masses $m_1$ and $m_2$ respectively, move on circular orbits around their common center of mass \citep{S67}. The third body, which plays the role of a test particle, moves inside the gravitational field created by the presence of the two primaries. This test particle has significantly smaller mass than the two primaries ($m \ll m_1$ and $m \ll m_2$) and therefore we can reasonably assume that it does not perturb or influence, in any way, the keplerian motion of the primaries.

A specific system of units (regarding length, mass and time) was adopted so that the gravitational constant $G$, the sum of the masses and the distance between the centers of the primaries to be equal to 1. For describing the motion of the third body we choose a rotating coordinate frame of reference where its origin is at the center of mass of the two primaries. The dimensionless masses of the primaries are $1-\mu$ and $\mu$, where $\mu = m_2/(m_1 + m_2) \leqslant 1/2$ is the mass ratio. Both primary bodies have their centers on the $x$-axis and specifically at $(x_1, 0)$ and $(x_2, 0)$, where $x_1 = -\mu$ and $x_2 = 1- \mu$.

We shall consider the general case of the photogravitational restricted three-body problem with oblateness. In particular, we assume that one of the primaries is an oblate spheroid which emits radiation. The second primary on the other hand it is assumed to be a non-radiating spherically symmetric body. This choice (two primaries with clear and distinct physical differences) will allow us to determine the influence of the oblateness and the radiation pressure on the Newton-Raphson basins of attraction.

The time-independent effective potential function of the photogravitational restricted three-body problem with oblateness, according to \citet{SSR75} and \citet{S82}, is
\begin{equation}
\Omega(x,y) = \frac{q\left(1 - \mu \right)}{r_1}\left(1 + \frac{A_1}{2r_1^2}\right) + \frac{\mu}{r_2} + \frac{n^2}{2}\left(x^2 + y^2 \right),
\label{pot}
\end{equation}
where $(x,y)$ are the coordinates of the test particle, while
\[
r_1 = \sqrt{\left(x - x_1\right)^2 + y^2},
\]
\begin{equation}
r_2 = \sqrt{\left(x - x_2\right)^2 + y^2},
\end{equation}
are the distances of the test particle from the oblate or radiating primary and spherically symmetric primary, respectively.

The angular velocity is
\begin{equation}
n = \sqrt{1 + \frac{3 A_1}{2}}.
\label{angv}
\end{equation}
The oblateness coefficient is defined as
\begin{equation}
A_1 = \frac{(RE)^2 - (RP)^2}{5R^2},
\label{obl0}
\end{equation}
where $R$ is the distance between the centers of the two primaries, while $RE$ and $RP$ are the equatorial and polar radius, respectively of the oblate primary. The radiation pressure is controlled through the parameter $q$ which is given by
\begin{equation}
q = 1 - \frac{F_p}{F_g},
\label{rad0}
\end{equation}
where $F_p$ is the solar radiation pressure force and $F_g$ is the gravitational attraction force \citep{SMB85}. In this work we shall restrict our investigation regarding the values of the oblateness and the radiation pressure factor in the intervals $A_1 \in [0, 0.5]$ and $q \in (0, 1]$ \citep[e.g.,][]{KDP06,PK06,Z15a,Z15b,Z15c}.

The equations which govern the motion of the test particle in the corotating frame of reference read
\[
\Omega_x = \frac{\partial \Omega}{\partial x} = \ddot{x} - 2n\dot{y},
\]
\begin{equation}
\Omega_y = \frac{\partial \Omega}{\partial y} = \ddot{y} + 2n\dot{x}.
\label{eqmot}
\end{equation}

The system of differential equations (\ref{eqmot}) admits only one integral of motion (known also as the Jacobi integral of motion). The corresponding Hamiltonian is
\begin{equation}
J(x,y,\dot{x},\dot{y}) = 2\Omega(x,y) - \left(\dot{x}^2 + \dot{y}^2 \right) = C,
\label{ham}
\end{equation}
where $\dot{x}$ and $\dot{y}$ are the velocities, while $C$ is the Jacobi constant which is conserved.

\section{Evolution of the equilibrium points}
\label{lgevol}

\begin{figure}[!tH]
\centering
\includegraphics[width=\hsize]{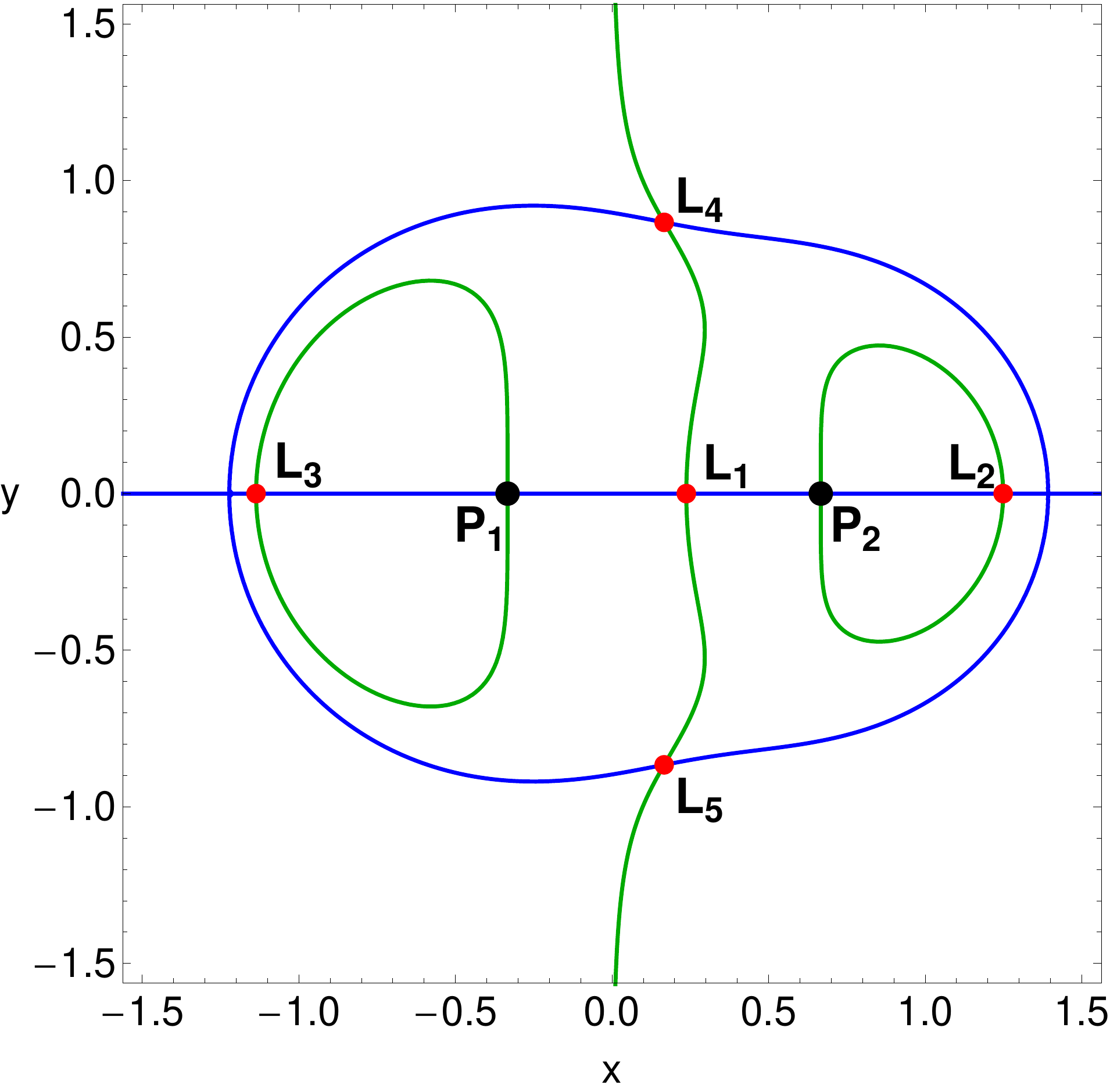}
\caption{The red dots indicate the positions of the five equilibrium points through the intersections of $\Omega_x = 0$ (green) and $\Omega_y = 0$ (blue), when $\mu = 1/3$, $A_1 = 0$, and $q = 1$. The black dots denote the centers of the two primaries.}
\label{lgs}
\end{figure}

It was found by Lagrange that five distinct three-body formations exist for two bodies which move in circular orbits around their common center of mass. For an observer in the rotating frame of reference these formations appear to be invariant. Moreover, these special five positions of the test particle for which its location appears to be stationary when viewed from the rotating frame of reference are called Lagrange libration points $L_i$, $i = 1, ..., 5$ \citep{S67}. The exact positions of the Lagrange points are the solutions of the system
\begin{align}
&\ddot{x} = \ddot{y} = \dot{x} = \dot{y} = 0, \nonumber\\
&\Omega_x = \Omega_y = 0.
\label{lps}
\end{align}
In Fig. \ref{lgs} we see how the intersections of Eqs. $\Omega_x = \Omega_y = 0$ define the positions of the equilibrium points when $\mu = 1/3$, $A_1 = 0$, and $q = 1$, that is the case of no oblateness and no radiation. Three of the equilibrium points, $L_1$, $L_2$, and $L_3$, (known as collinear points) are located on the $x$-axis, while the other two $L_4$ and $L_5$ are called triangular points and they are located on the vertices of equilateral triangles. At this point, it should be emphasized that the labeling of the collinear points is not consistent throughout the literature. In this paper, we adopt the most popular case according to which $L_1$ lies between the two primary bodies, $L_2$ is at the right side of $P_2$ (the spherically symmetric primary), while $L_3$ is at the left side of $P_1$ (the oblate or radiating primary). Therefore we have
\begin{equation}
x_{L_3} < x_1 < x_{L_1} < x_2 < x_{L_2}.
\label{pos}
\end{equation}

\begin{figure*}[!tH]
\centering
\resizebox{\hsize}{!}{\includegraphics{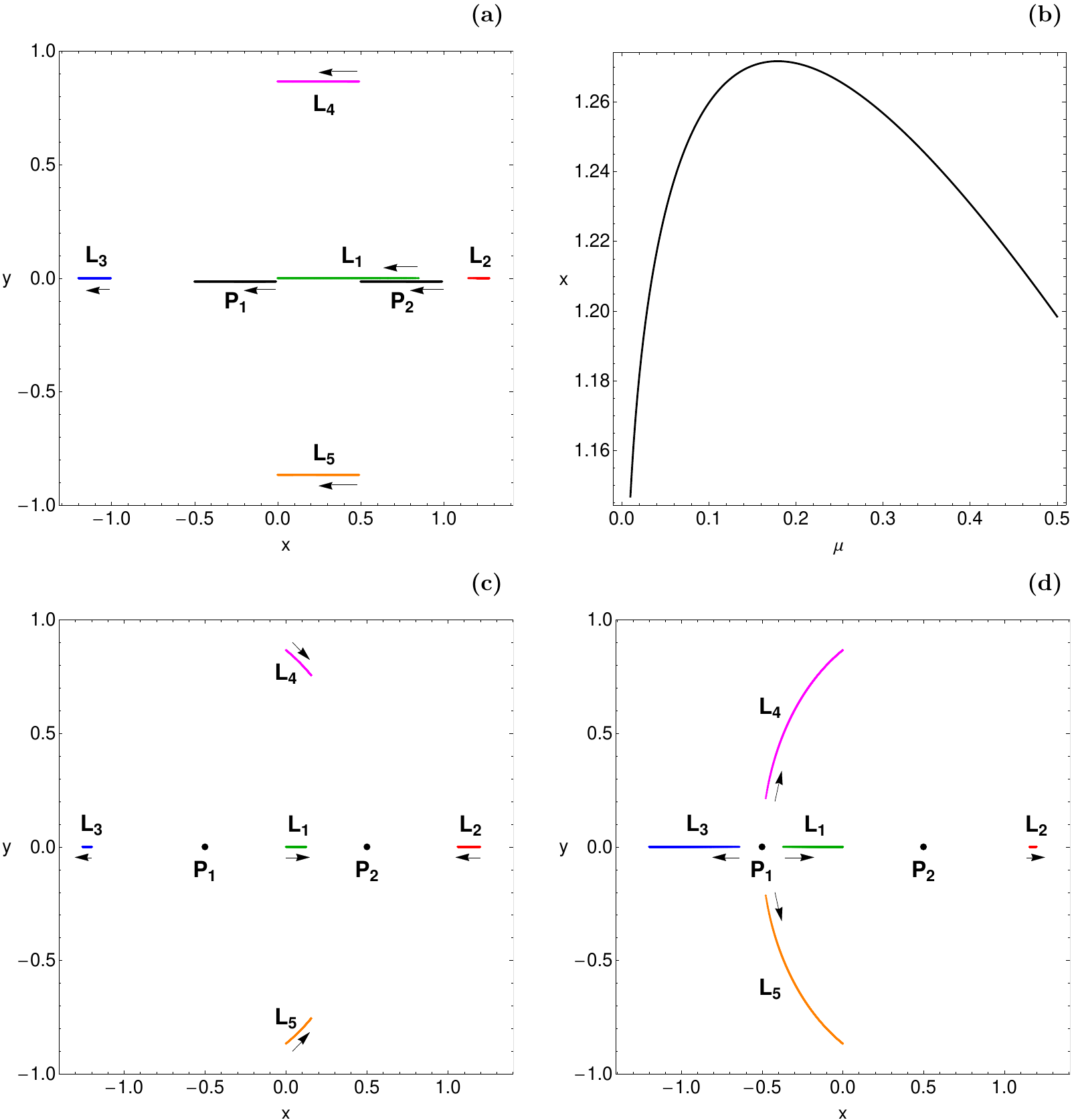}}
\caption{The evolution of the equilibrium points in the planar circular restricted three-body problem with oblateness and radiation pressure when (a-upper left): $\mu \in [0.01, 0.5]$, $A_1 = 0$, and $q = 1$, (c-lower left): $\mu = 1/2$, $A_1 \in [0, 0.5]$, and $q = 1$, and (d-lower right): $\mu = 1/2$, $A_1 = 0$, and $q \in [0.01, 1]$. (b-upper right): The evolution of the $x$ coordinate of $L_2$ with variable mass ratio. The arrows indicate the movement direction of the equilibrium points as the variable parameter in each case increases.}
\label{lgse}
\end{figure*}

The stability of the equilibrium points, where oblateness and radiation pressure are present, has been investigated in many previous works \citep[e.g.,][]{D11,PKPN15,STB01,SL12}. In this paper, we shall explore how the mass ratio $(\mu)$, the oblateness coefficient $(A_1)$, and radiation pressure factor $(q)$ influence the positions of the libration points. Our results are illustrated in Fig. \ref{lgse}(a-d). In panel (a) we see the space-evolution of the five equilibrium points as well as of the centers of the primaries when $\mu \in [0.01, 0.5]$, $A_1 = 0$, and $q = 1$, that is the case of no oblateness and no radiation pressure. One may observe that all libration points, except $L_2$, and the centers of the primaries are moved to the left. It should be emphasized that for the triangular points $L_4$ and $L_5$ only the $x$ coordinate changes, while the $y$ coordinate remains constant at $\pm \sqrt{3}/2$. The space-evolution of $x_{L_2}$ on the other hand, is not monotonic as it increases when $0.01 \leq \mu < 0.179$, while it decreases when $0.179 < \mu \leq 0.5$ (see panel (b) of Fig. \ref{lgse}). In the same vein, in panel (c) we present the space-evolution of the equilibrium points when $\mu = 1/2$, $A_1 \in [0, 0.5]$, and $q = 1$, that is the case of primaries with equal masses and no radiation. In this case, $L_1$ and $L_3$ move away from the center $P_1$, while $L_2$ approaches the center $P_2$, as the left primary becomes more and more oblate. Moreover, now both coordinates of the triangular points change with increasing $A_1$. Finally in panel (d) of Fig. \ref{lgse} we depict the space-evolution of the libration points when $\mu = 1/2$, $A_1 = 0$, and $q \in [0.01,1]$, that is the case of spherically symmetric (no oblateness) primaries with equal masses. Here the changes regarding the positions of the four of the five equilibrium points are much more prominent with respect to the previous cases (variable mass ratio and variable oblateness). Our numerical calculations suggest that in the limiting case where $q \rightarrow 0$ $L_1$, $L_3$, $L_4$, and $L_5$ tend to collide with the center $P_1$. Taking into consideration the above-mentioned analysis we may conclude that the influence of the mass ratio $\mu$ and the radiation pressure factor $q$ on the positions of the equilibrium points is much more stronger than that of the oblateness coefficient $A_1$. Here we would like to note that we did not consider cases with prolate primaries (with negative values of $A_1$).

\section{The Newton-Raphson basins of attraction}
\label{bas}

We decided to use the multivariate version of the Newton-Raphson method, a simple yet a very accurate computational tool, in order to determine to which of the five equilibrium points each initial point on the configuration $(x,y)$ plane leads to. The Newton-Raphson method is applicable to systems of multivariate functions $f({\bf{x}}) = 0$, through the iterative scheme \begin{equation}
{\bf{x}}_{n+1} = {\bf{x}}_{n} - J^{-1}f({\bf{x}}_{n}),
\label{iter}
\end{equation}
where $J^{-1}$ is the inverse Jacobian matrix of $f({\bf{x_n}})$. In our case the system of equations is
\begin{equation}
\begin{cases}
\Omega_x = 0 \\
\Omega_y = 0.
\end{cases}
\label{sys}
\end{equation}
After trivial calculations the iterative formulae take the form
\begin{eqnarray}
x_{n+1} &=& x_n - \left( \frac{\Omega_x \Omega_{yy} - \Omega_y \Omega_{xy}}{\Omega_{yy} \Omega_{xx} - \Omega^2_{xy}} \right)_{(x_n,y_n)}, \nonumber\\
y_{n+1} &=& y_n + \left( \frac{\Omega_x \Omega_{yx} - \Omega_y \Omega_{xx}}{\Omega_{yy} \Omega_{xx} - \Omega^2_{xy}} \right)_{(x_n,y_n)},
\label{nrm}
\end{eqnarray}
where $x_n$, $y_n$ are the values of the $x$ and $y$ variables at the $n$-th step of the iterative process, while the subscripts of $\Omega$ denote the corresponding partial derivatives of the potential function. The multivariate Newton-Raphson method has also been used to obtain the basins of attraction in other dynamical systems, such as the restricted three-body problem \citep[e.g.,][]{KGK12}, the four-body problem \citep[e.g.,][]{BP11,KK14}, or even the ring problem of $N + 1$ bodies \citep[e.g.,][]{CK07}.

The Newton-Raphson algorithm is activated when an initial condition $(x_0,y_0)$ on the configuration plane is given, while it stops when the positions of the equilibrium points are reached, with some predefined accuracy. All the initial conditions that lead to a specific equilibrium point, compose a basin of attraction or an attracting region. Here we would like to clarify that the Newton-Raphson basins of attraction should not be mistaken with the classical basins of attraction in dissipative systems. We observe that the iterative formulae (\ref{nrm}) include both the first and the second derivatives of the effective potential function $\Omega(x,y)$ and therefore we may claim that the obtained numerical results directly reflect some of the basic qualitative characteristics of the dynamical system. The major advantage of knowing the Newton-Raphson basins of attraction in a dynamical system is the fact that we can select the most favorable initial conditions, with respect to required computation time, when searching for an equilibrium point.

For obtaining the basins of convergence we worked as follows: First we defined a dense uniform grid of $1024 \times 1024$ initial conditions regularly distributed on the configuration $(x,y)$ space. The iterative process was terminated when an accuracy of $10^{-15}$ has been reached, while we classified all the $(x,y)$ initial conditions that lead to a particular solution (equilibrium point). At the same time, for each initial point, we recorded the number $(N)$ of iterations required to obtain the aforementioned accuracy. Logically, the required number of iterations for locating an equilibrium point strongly depends on the value of the predefined accuracy. All the computations reported in this paper were performed using a double precision \verb!FORTRAN 77! algorithm. Furthermore, all graphics have been created using the version 10.3 of Mathematica$^{\circledR}$ \citep{W03}.

In the following we shall try to determine how the dynamical quantities $\mu$, $A_1$ and $q$ influence the Newton-Raphson basins of attraction. In order to focus to the influence of each dynamical quantity we will examine them separately.

\subsection{The influence of the mass ratio $\mu$}
\label{ss1}

\begin{figure*}[!tH]
\centering
\resizebox{0.85\hsize}{!}{\includegraphics{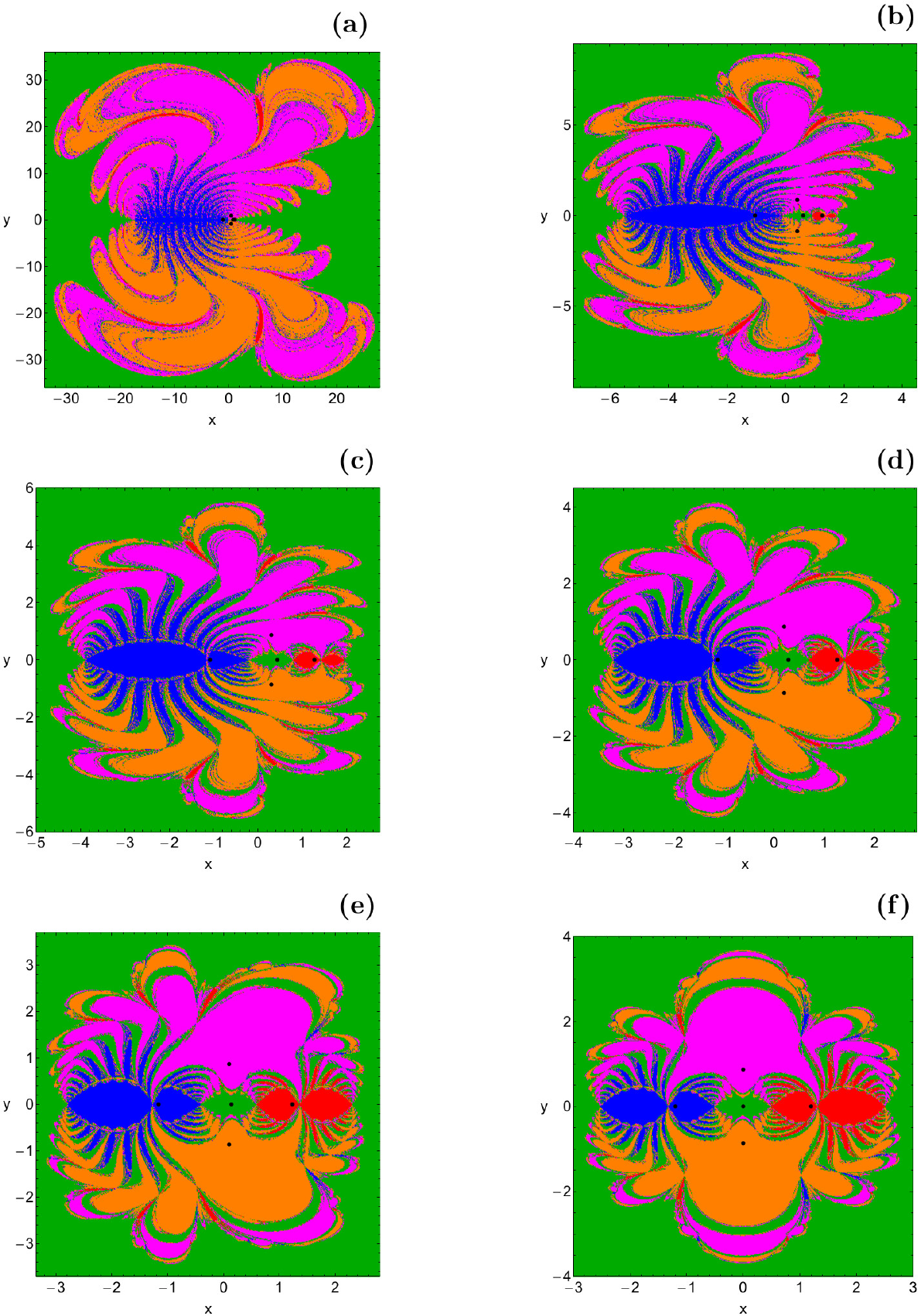}}
\caption{The Newton-Raphson basins of attraction on the configuration $(x,y)$ plane when $\mu$ varies in the interval $[0.01, 0.5]$, while $A_1 = 0$ and $q = 1$. The positions of the five equilibrium points are indicated by black dots. (a): $\mu = 0.01$; (b): $\mu = 0.1$; (c): $\mu = 0.2$; (d): $\mu = 0.3$; (e): $\mu = 0.4$; (f): $\mu = 0.5$. The color code denoting the attractors is as follows: $L_1$ (green); $L_2$ (red); $L_3$ (blue); $L_4$ (magenta); $L_5$ (orange); non-converging points (white).}
\label{mass}
\end{figure*}

\begin{figure*}[!tH]
\centering
\resizebox{0.95\hsize}{!}{\includegraphics{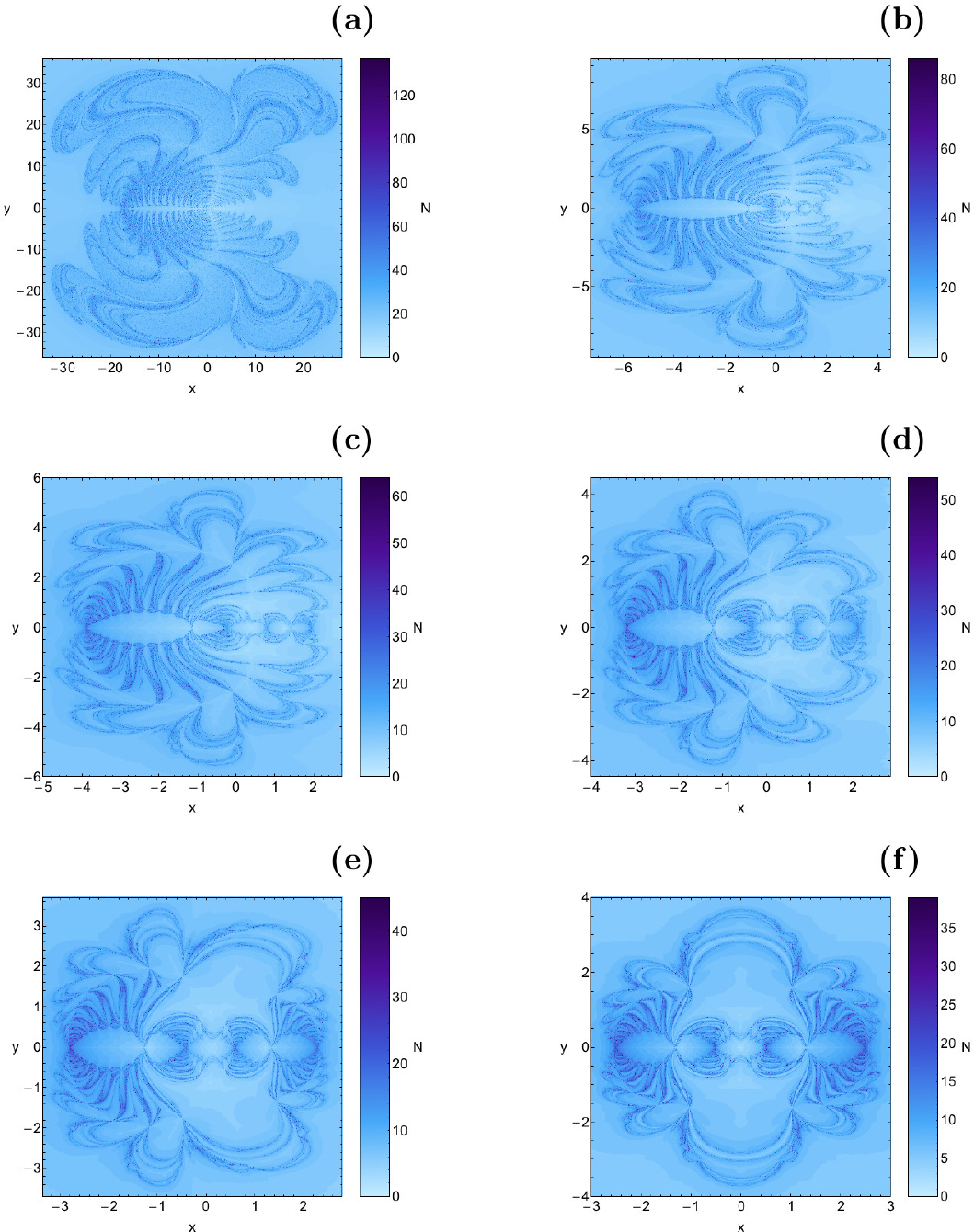}}
\caption{The distribution of the corresponding number $(N)$ of required iterations for obtaining the Newton-Raphson basins of attraction shown in Fig. \ref{mass}(a-f).}
\label{massn}
\end{figure*}

\begin{figure*}[!tH]
\centering
\resizebox{0.85\hsize}{!}{\includegraphics{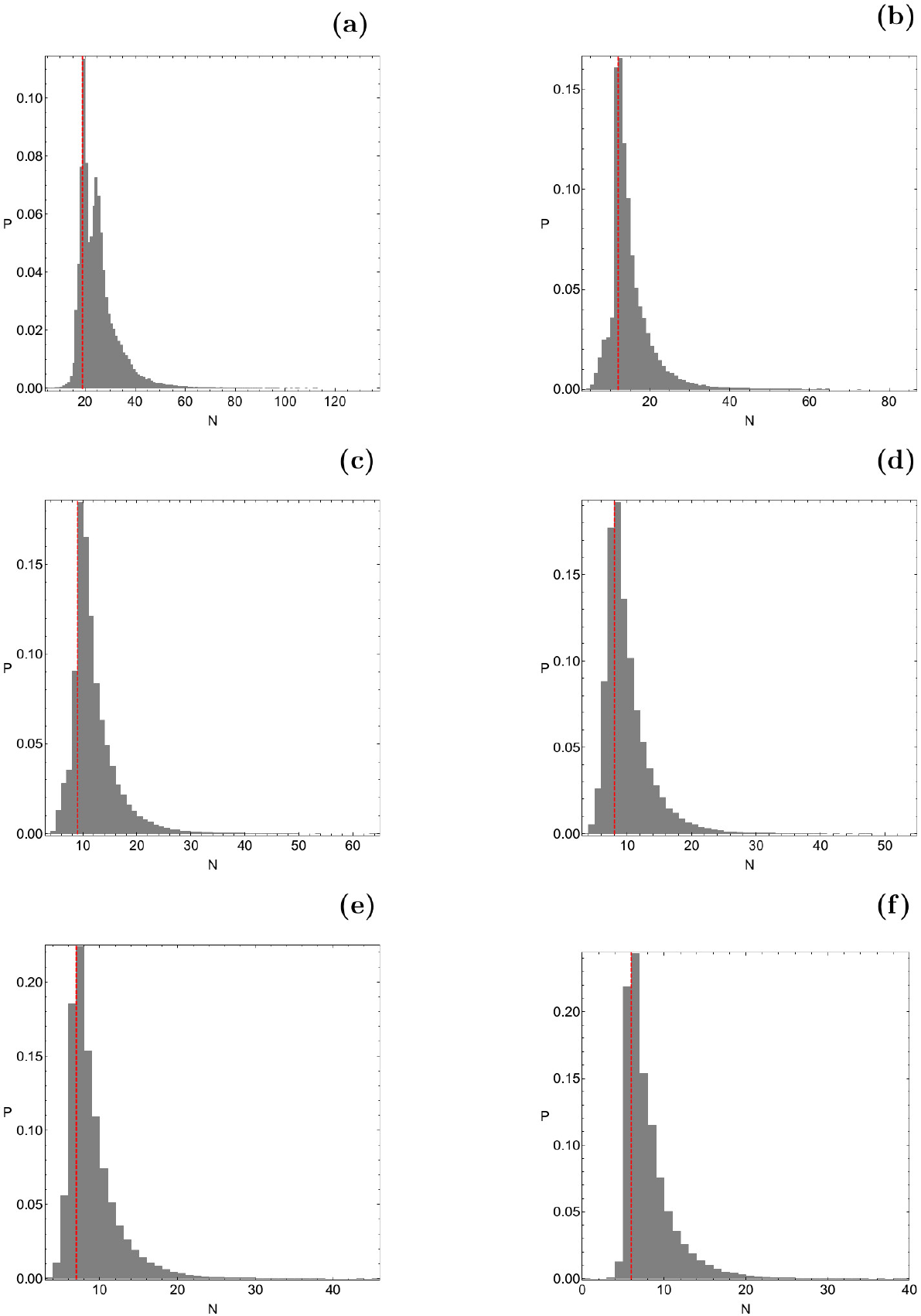}}
\caption{The corresponding probability distribution of required iterations for obtaining the Newton-Raphson basins of attraction shown in Fig. \ref{mass}(a-f). The vertical, dashed, red line indicates, in each case, the most probable number $(N^{*})$ of iterations.}
\label{massp}
\end{figure*}

\begin{figure*}[!tH]
\centering
\resizebox{0.85\hsize}{!}{\includegraphics{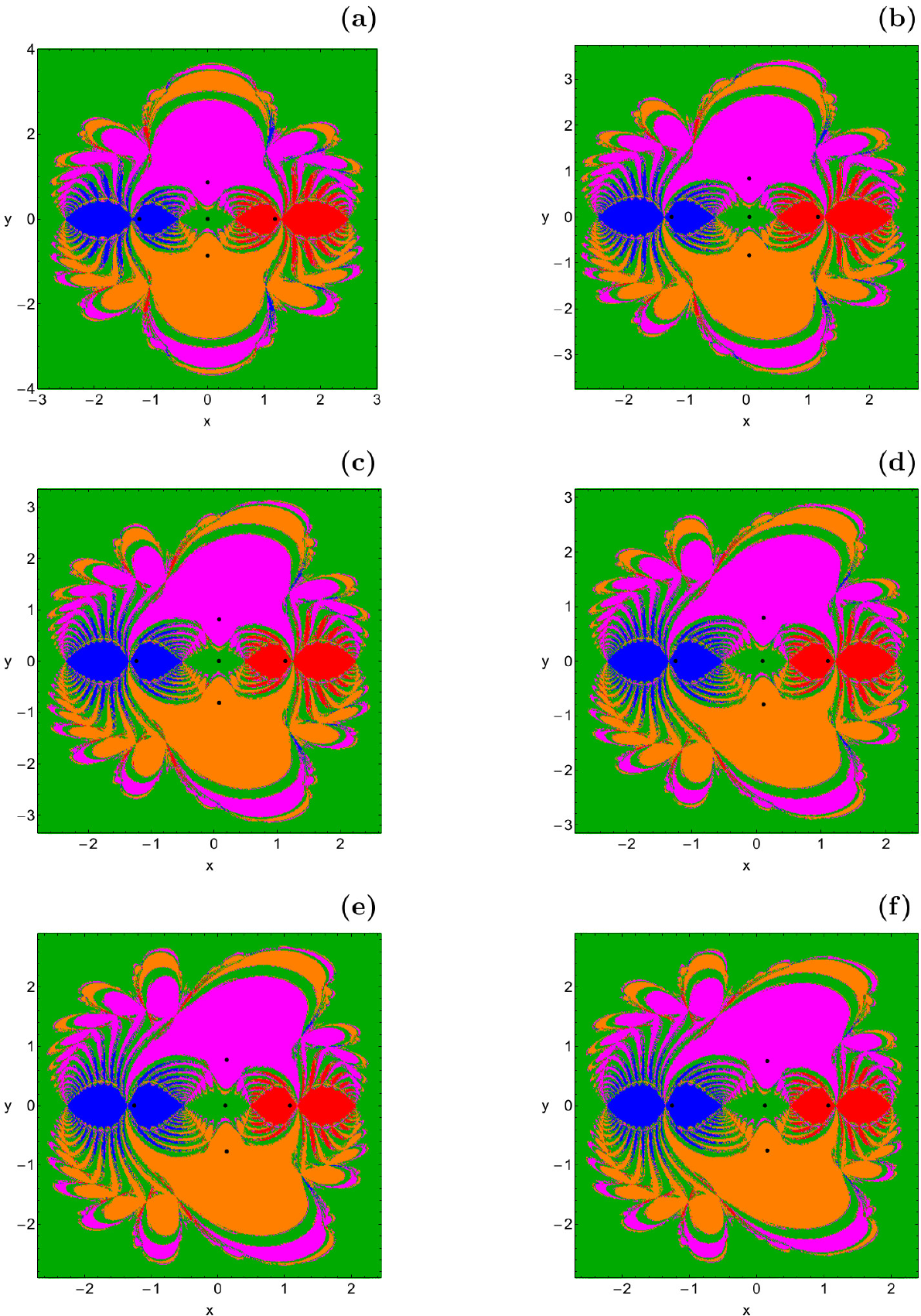}}
\caption{The Newton-Raphson basins of attraction on the configuration $(x,y)$ plane when $A_1$ varies in the interval $[0, 0.5]$, while $\mu = 1/2$ and $q = 1$. (a): $A_1 = 0.01$; (b): $A_1 = 0.1$; (c): $A_1 = 0.2$; (d): $A_1 = 0.3$; (e): $A_1 = 0.4$; (f): $A_1 = 0.5$. The color code denoting the attractors is as in Fig. \ref{mass}.}
\label{obl}
\end{figure*}

\begin{figure*}[!tH]
\centering
\resizebox{0.95\hsize}{!}{\includegraphics{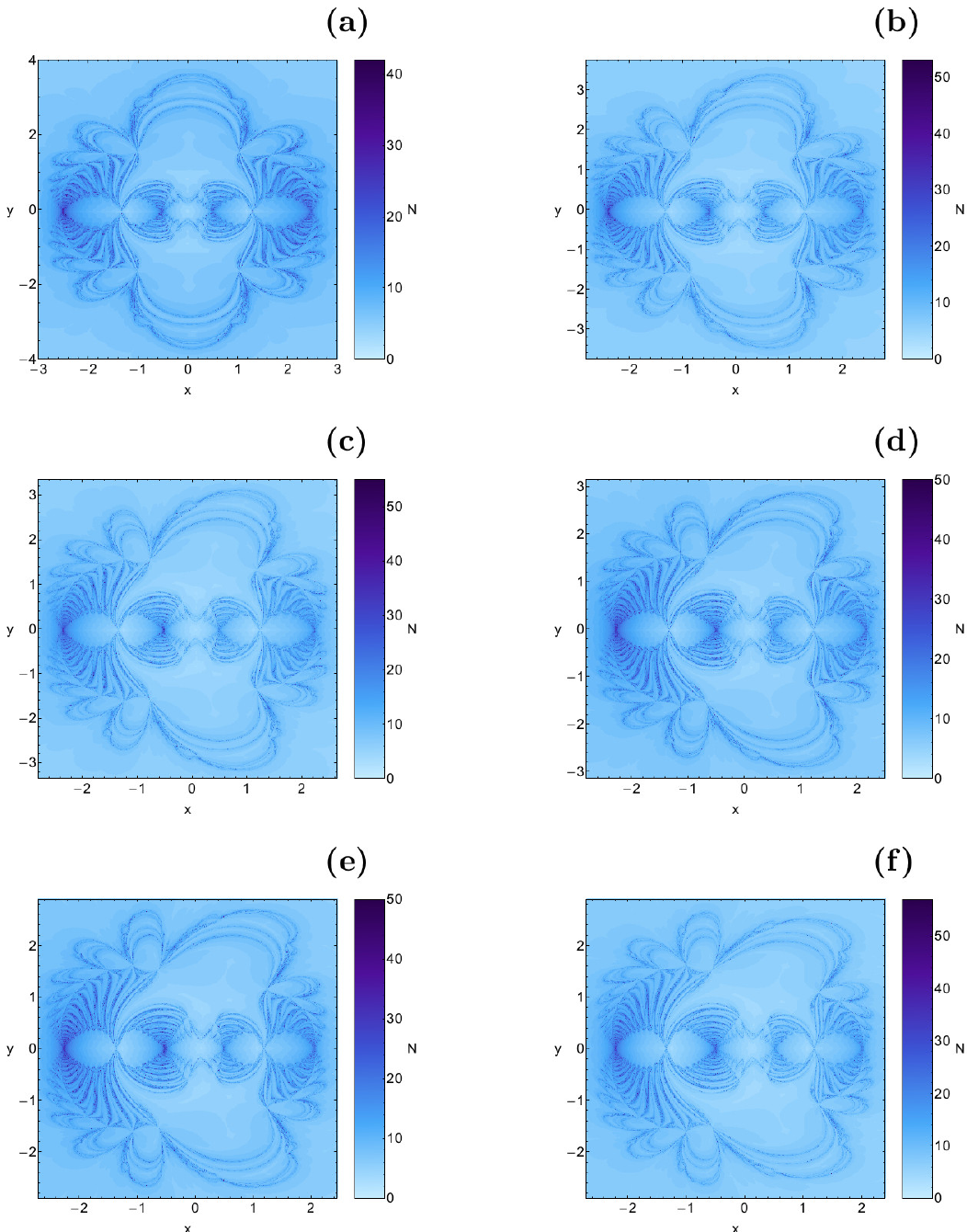}}
\caption{The distribution of the corresponding number $(N)$ of required iterations for obtaining the Newton-Raphson basins of attraction shown in Fig. \ref{obl}(a-f).}
\label{obln}
\end{figure*}

\begin{figure*}[!tH]
\centering
\resizebox{0.85\hsize}{!}{\includegraphics{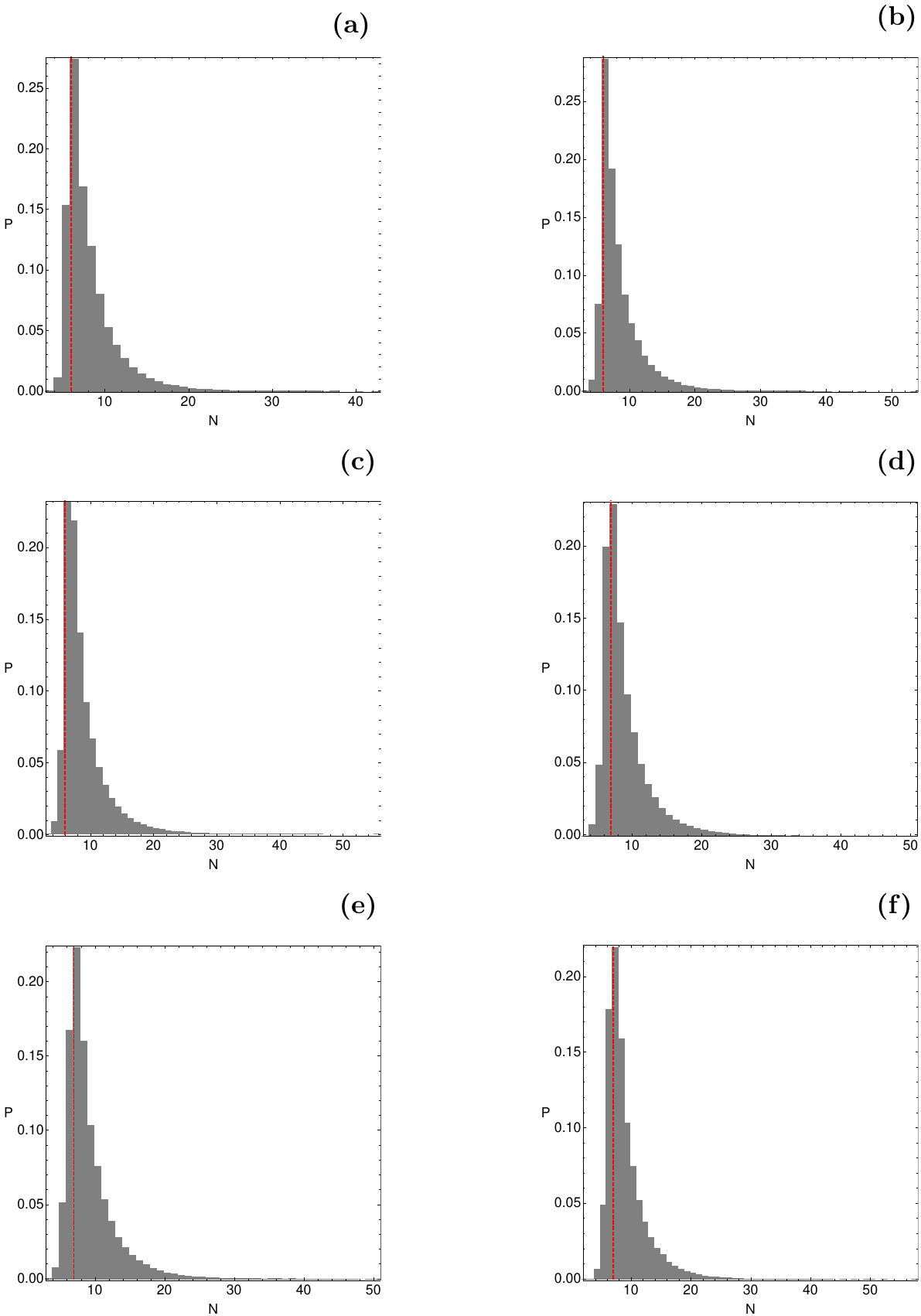}}
\caption{The corresponding probability distribution of required iterations for obtaining the Newton-Raphson basins of attraction shown in Fig. \ref{obl}(a-f). The vertical, dashed, red line indicates, in each case, the most probable number $(N^{*})$ of iterations.}
\label{oblp}
\end{figure*}

\begin{figure*}[!tH]
\centering
\resizebox{0.85\hsize}{!}{\includegraphics{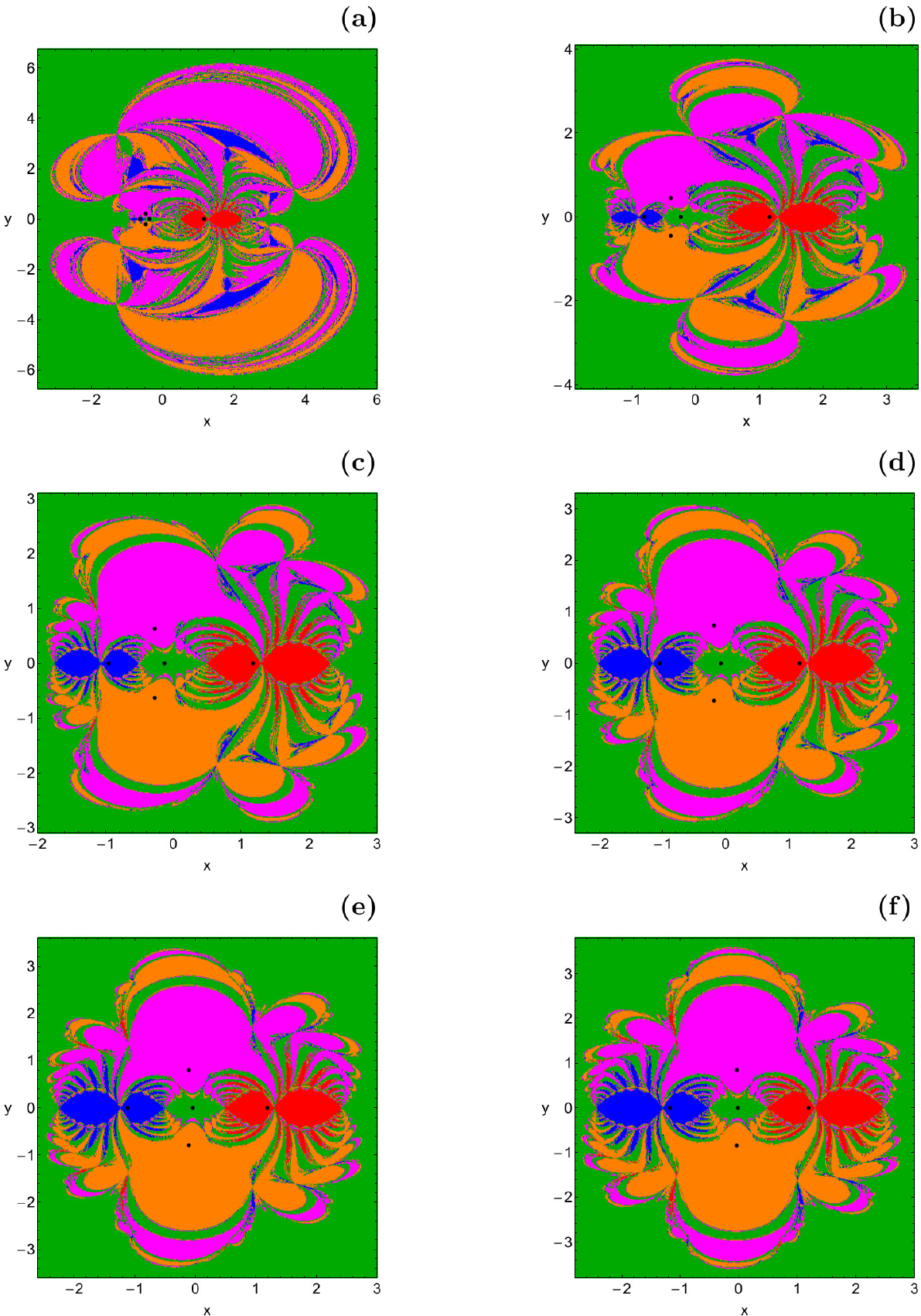}}
\caption{The Newton-Raphson basins of attraction on the configuration $(x,y)$ plane when $q$ varies in the interval $[0.01, 1]$, while $\mu = 1/2$ and $A_1 = 0$. (a): $q = 0.01$; (b): $q = 0.1$; (c): $q = 0.3$; (d): $q = 0.5$; (e): $q = 0.7$; (f): $q = 0.9$. The color code denoting the attractors is as in Fig. \ref{mass}.}
\label{rad}
\end{figure*}

\begin{figure*}[!tH]
\centering
\resizebox{0.95\hsize}{!}{\includegraphics{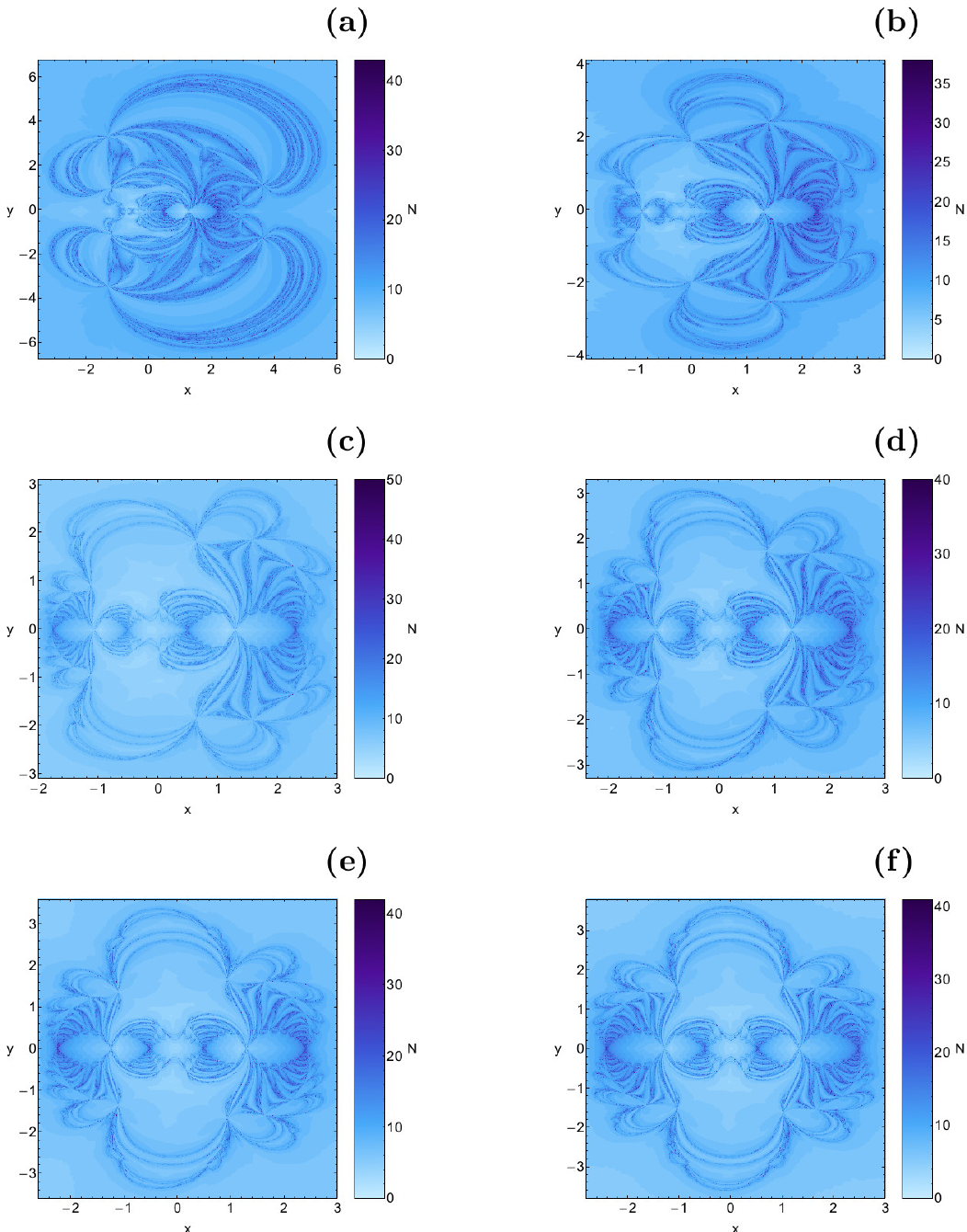}}
\caption{The distribution of the corresponding number $(N)$ of required iterations for obtaining the Newton-Raphson basins of attraction shown in Fig. \ref{rad}(a-f).}
\label{radn}
\end{figure*}

\begin{figure*}[!tH]
\centering
\resizebox{0.85\hsize}{!}{\includegraphics{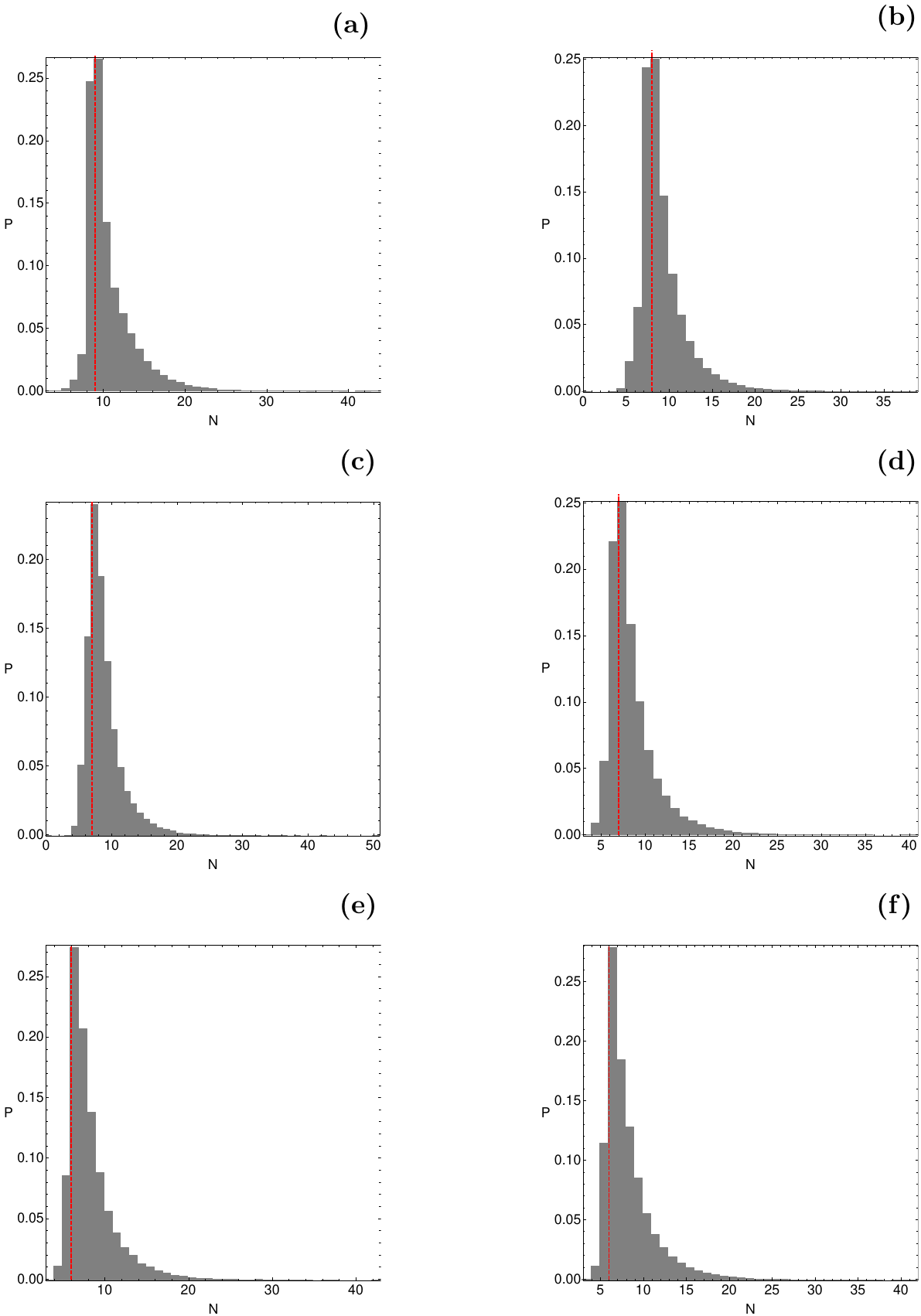}}
\caption{The corresponding probability distribution of required iterations for obtaining the Newton-Raphson basins of attraction shown in Fig. \ref{rad}(a-f). The vertical, dashed, red line indicates, in each case, the most probable number $(N^{*})$ of iterations.}
\label{radp}
\end{figure*}

\begin{figure*}[!tH]
\centering
\resizebox{\hsize}{!}{\includegraphics{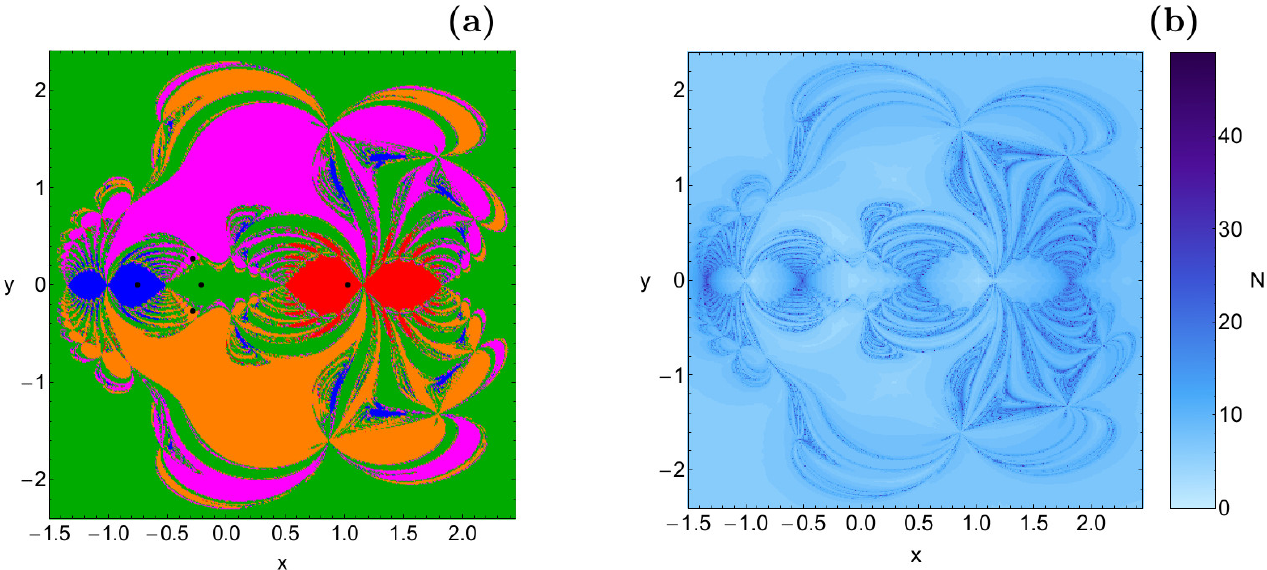}}
\caption{(a-left): The Newton-Raphson basins of attraction on the configuration $(x,y)$ plane for the extreme case where $\mu = 1/2$, $A_1 = 0.5$, and $q = 0.01$. The color code denoting the attractors is as in Fig. \ref{mass}. (b-right): The distribution of the corresponding number $(N)$ of required iterations for obtaining the basins of attraction of panel (a).}
\label{extr}
\end{figure*}

\begin{figure}[!tH]
\centering
\includegraphics[width=\hsize]{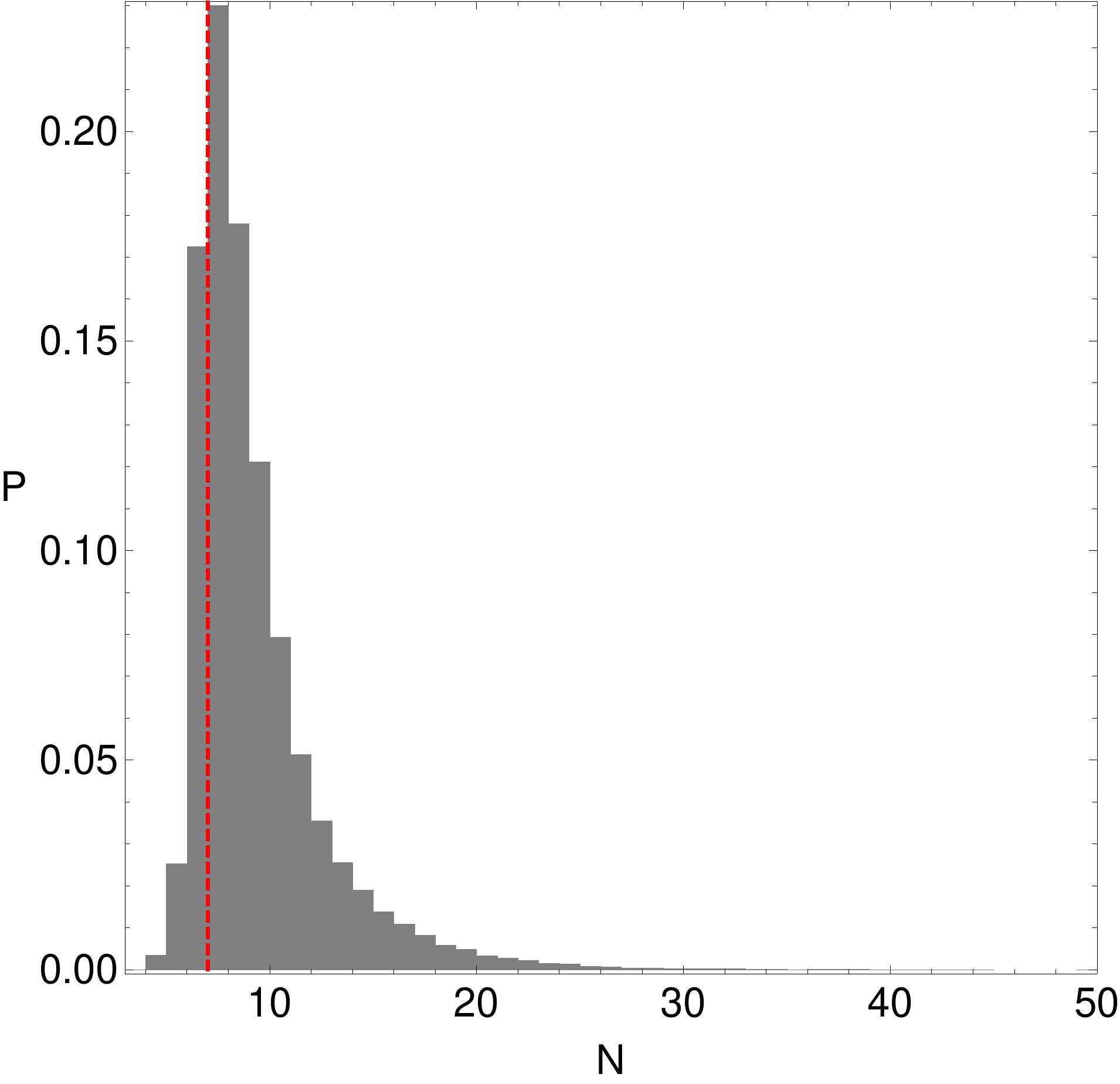}
\caption{The corresponding probability distribution of required iterations for obtaining the Newton-Raphson basins of attraction shown in Fig. \ref{extr}. The vertical, dashed, red line indicates, the most probable number $(N^{*})$ of iterations.}
\label{hist}
\end{figure}

Our investigation begins with the case where the mass ratio $\mu$ varies in the interval $[0.01, 0.5]$, while $A_1 = 0$ and $q = 1$. In other words when both primaries are spherically symmetric, while there is no radiation pressure. In Fig. \ref{mass}(a-f) we present the Newton-Raphson basins of attraction for six values of the mass ratio. Looking the diagrams we may say that the shape of the basins of attraction corresponding to equilibrium points $L_2$ and $L_3$ have the shape of bugs with many legs and many antennas. Furthermore, the shape of the basins of attraction corresponding to the triangular equilibrium points $L_4$ and $L_5$ looks like multiple butterfly wings. It is interesting to note that the basins of attraction corresponding to the central $L_1$ extend to infinity, while on the other hand the area of all the other basins of convergence is finite. It is evident that a large portion of the configuration $(x,y)$ plane is covered by well-formed basins of attraction. The boundaries between the several basins of convergence however are highly fractal\footnote{When we state that an area is fractal we simply mean that it has a fractal-like geometry without conducting any specific calculations for computing the fractal dimensions.} and they look like a ``chaotic sea". This means that if we choose a starting point $(x_0,y_0)$ of the Newton-Raphson method inside these fractal domains we will observe that our choice is very sensitive. In particular, a slight change in the initial conditions leads to completely different final destination (different attractor) and therefore the beforehand prediction becomes extremely difficult. In Fig. \ref{massn}(a-f) we provide the distribution of the corresponding number $(N)$ of iterations required for obtaining the desired accuracy, using tones of blue. In the same vein, in Fig. \ref{massp}(a-f) the corresponding probability distribution of iterations is shown. The probability $P$ is defined as follows: let's assume that $N_0$ initial conditions $(x_0,y_0)$ converge to one of the five attractors after $N$ iterations. Then $P = N_0/N_t$, where $N_t$ is the total number of initial conditions in every grid.

Combining all the information form Figs. \ref{mass}, \ref{massn}, and \ref{massp} we conclude that the most important phenomena which take place as the value of the mass ratio increases are the following:
\begin{itemize}
  \item The area of the basins of attraction corresponding to equilibrium point $L_3$ decreases, while at the same time the area of the basins of attraction corresponding to equilibrium point $L_2$ increases. For $\mu > 0.1$ the area of the basins of attraction corresponding to the central libration point $L_1$ as well as to the triangular points $L_4$ and $L_5$ remains almost unperturbed.
  \item The required number $(N)$ of iterations for obtaining the desired accuracy constantly decreases. Consequently, the most probable number $(N^{*})$ of iterations also decreases. In all examined cases, for more than 95\% of the initial conditions on the configuration $(x,y)$ plane the iterative formulae (\ref{nrm}) need no more than 50 iterations for obtaining the desired accuracy.
\end{itemize}

\subsection{The influence of the oblateness coefficient $A_1$}
\label{ss2}

We continue with the case where the oblateness coefficient $A_1$ varies in the interval $[0.01, 0.5]$, while $\mu = 1/2$ and $q = 1$. In other words when both primaries (the oblate spheroid and the spherically symmetric) have the same mass, while there is no radiation pressure. We decided to have primaries with equal masses (Copenhagen problem) so as the shape of them to be the only difference. The Newton-Raphson basins of attraction for six values of the oblateness coefficient are presented in Fig. \ref{obl}(a-f), while Fig. \ref{obln}(a-f) shows the distribution of the corresponding number $(N)$ of iterations required for obtaining the desired accuracy. The corresponding probability distribution of iterations is given in Fig. \ref{oblp}(a-f).

Taking into consideration all the numerical outcomes presented in Figs. \ref{obl}, \ref{obln}, and \ref{oblp} we could argue that the most important phenomena which take place as the left primary body becomes more oblate are the following:
\begin{itemize}
  \item The area on the configuration $(x,y)$ plane corresponding to the central equilibrium point $L_1$ as well as to the triangular Lagrange points $L_4$ and $L_5$ remains almost unperturbed. For $A_1 > 0.2$ the area corresponding to equilibrium point $L_2$ seems to reduce. On the other hand, the area occupied by basins of attraction corresponding to Lagrange point $L_3$ is constantly increases throughout the interval [0.01, 0.5]. This is true because with a closer look to the plots of Fig. \ref{obl}(a-f) it is seen that the number of legs and antennas of the left bug increases.
  \item The required number $(N)$ of iterations for obtaining the desired accuracy fluctuates without showing any clear sings of decrease or increase. Therefore, the most probable number $(N^{*})$ of iterations remains almost constant and it is equal to 6 or 7. In all examined cases, for more than 95\% of the initial conditions on the configuration $(x,y)$ plane the iterative formulae (\ref{nrm}) need no more than 40 iterations for obtaining the desired accuracy.
\end{itemize}

\subsection{The influence of the radiation pressure factor $q$}
\label{ss3}

The last case under consideration follows the scenario according to which the radiation pressure factor $q$ varies in the interval $[0.01, 0.9]$, while $\mu = 1/2$ and $A_1 = 0$. This means that both primaries (radiating or not) have the same mass, while both of them are spherically symmetric. Our choice to have primaries with equal masses and identical shape is justified if we take into account that now the only difference between the two primaries is the presence of radiation pressure and therefore we can focus to its influence on the basins of convergence. In the following Fig. \ref{obl}(a-f) we illustrate the Newton-Raphson basins of attraction for six values of the radiation pressure factor, while in Fig. \ref{radn}(a-f) we present the distribution of the corresponding number $(N)$ of iterations required for obtaining the desired accuracy. Finally Fig. \ref{radp}(a-f) shows the corresponding probability distribution of iterations.

Correlating all the numerical results given in Figs. \ref{rad}, \ref{radn}, and \ref{radp} one may reasonably deduce that the most important phenomena which take place as the the radiation pressure of the left primary body decreases (remember that the intensity of the radiation pressure decreases as the value of $q$ approaches to 1) are the following:
\begin{itemize}
  \item The area on the configuration $(x,y)$ plane covered by basins of attraction corresponding to the central equilibrium point $L_1$ and to the triangular points $L_4$ and $L_5$ remains almost unperturbed. On the contrary, the area corresponding to the Lagrange point $L_2$ increases, while for $q > 0.5$ it seems to saturate. The area covered by initial conditions which converge to the equilibrium point $L_3$ exhibits a constant increase throughout the interval [0.01, 0.9]. It is interesting to note that in this case the shape of the basins of attraction of the triangular points is highly affected by the radiation pressure factor, even though the corresponding percentages of the basins do not change.
  \item The average value of required number $(N)$ of iterations for obtaining the desired accuracy decreases. Consequently, the the most probable number $(N^{*})$ of iterations is reduced from 9 when $q = 0.01$ to 6 when $q = 0.9$. In all examined cases, for more than 95\% of the initial conditions on the configuration $(x,y)$ plane the iterative formulae (\ref{nrm}) need no more than 30 iterations for obtaining the desired accuracy.
\end{itemize}

Before closing this section we would like to present the extreme case where the primary located at $P_1$ is highly oblate, while at the same time it is an intense emitter of radiation. In Fig. \ref{extr}a we can observe the Newton-Raphson basins of attraction when $\mu = 1/2$, $A_1 = 0.5$, and $q = 0.01$. For both the oblateness and the radiation pressure we adopted the highest possible values. The distribution of the corresponding number $(N)$ of iterations required for obtaining the desired accuracy is shown in Fig. \ref{extr}b. Our calculations suggest that in this case the majority, about 23\% (see Fig. \ref{hist}), of the initial conditions need 7 iterations in order to converge to one of the five attractors of the system.

\section{Discussion and conclusions}
\label{disc}

The aim of this paper was to obtain the basins of attraction in the planar circular restricted three-body problem where one of the primary bodies in an oblate spheroid or emitter of radiation. The basins of convergence for the five equilibrium points of the dynamical system have been determined with the help of the multivariate version of the Newton-Raphson method. These basins describe how each point on the configuration $(x,y)$ plane is attracted by one of the five attractors. Our thorough and systematic numerical investigation revealed how the position of the equilibrium points and the structure of the basins of attraction are influenced by the several dynamical parameters (i.e., the mass ratio $\mu$, the oblateness coefficient $A_1$ and the radiation pressure factor $q$). We also found correlations between the basins of attraction and the distribution of the corresponding required number of iterations.

For the numerical calculations of the sets of the initial conditions on the configuration $(x,y)$ plane, we needed about 3 minutes of CPU time on a Quad-Core i7 2.4 GHz PC, depending of course on the required number of iterations. When an initial condition had reached one of the five attractors the iterative procedure was effectively ended and proceeded to the next available initial condition.

We obtained quantitative information regarding the Newton-Raphson basins of attraction in the restricted three-body problem with oblateness and radiation pressure. The main results of our numerical research can be summarized as follows:
\begin{enumerate}
  \item In all examined cases, the configuration $(x,y)$ plane is a complicated mixture of basins of attraction and highly fractal regions. These regions are the exact opposite of the basins of attraction and they are completely intertwined with respect to each other (fractal structure). This sensitivity towards slight changes in the initial conditions in the fractal regions implies that it is impossible to predict the final state.
  \item The several basins of attraction are very intricately interwoven and they appear either as well-defined broad regions or as thin elongated bands. The fractal domains are mainly located in the vicinity of the basin boundaries.
  \item The area of the basins of attractions corresponding to equilibrium points $L_2$, $L_3$ as well as to triangular points $L_4$ and $L_5$ is finite. Additional numerical computations reveal that the area of the basins of convergence corresponding to the central equilibrium point $L_1$ extends to infinity.
  \item Our calculations strongly suggest that all initial conditions on the configuration plane converge, sooner or later, to one of the five attractors of the dynamical system. In other words, we did not encounter any non-converging initial condition.
  \item The iterative method was found to converge relatively fast with initial conditions inside the basins of attraction. On the other hand, the highest numbers of required iterations correspond to initial conditions in the fractal domains.
  \item Our results indicate that the mass ratio and the radiation pressure factor are the most influential dynamical quantities. On the contrary, the changes observed on the configuration plane due to the increase of the oblateness coefficient are much more milder.
  \item The number of iterations for obtaining the required accuracy is reduced when the mass ratio increases and also when the radiation pressure decreases. According to our analysis, for the case where the oblateness of the left primary body increases there is no clear sign regarding the tendency of the number of iterations.
\end{enumerate}

Taking into account the detailed and novel outcomes of our numerical exploration we may suggest that our computational task has been successfully completed. We hope that the present numerical analysis and the corresponding results to be useful in the field of basins of attraction of equilibrium points. It is in our future plans to expand our investigation in three dimensions thus revealing the basins of attraction inside the $(x,y,z)$ space. Furthermore, it would be very interesting to use other iterative schemes of higher order than that of the Newton-Raphson and compare the similarities and differences regarding the structure of the basins of convergence.

\section*{Acknowledgments}

I would like to express my warmest thanks to the anonymous referee for the careful reading of the manuscript and for all the apt suggestions and comments which allowed us to improve both the quality and the clarity of the paper.

\end{document}